\documentclass[12pt]{article}
\usepackage{amsmath}
\usepackage{amssymb}
\usepackage{graphics}
\usepackage{times}
\begin{document}
\thispagestyle{empty}
\begin{center}
{\LARGE \textbf{Spacetime is spinorial; new dimensions are timelike}}

\vspace{10pt}
{\large George A.J. Sparling\\\vspace{4pt} Laboratory of Axiomatics \\University of Pittsburgh\\\vspace{6pt} Pittsburgh, Pennsylvania, 15260, USA\bigskip }
\vspace{-25pt}
\begin{quote}\textbf{
\noindent \\
\noindent
Since Pythagoras of Samos and Euclid of Alexandria$^1$, we have known how to express the squared distance between entities as the sum of squares of
displacements in perpendicular directions.  Since Hermann Minkowski$^2$ and Albert Einstein$^3$, the squared interval between events has acquired a new term which
represents the square of the time displacement and which comes in with
a negative sign.  Most higher dimensional theories, whose aim is to unify the
physical interactions of nature, use space-like extra dimensions (more squares with positive signs)  rather than  time-like (more squares with negative signs)$^{[4-8]}$.  But there need be no contradiction if  timelike extra dimensions are used:  for example, in the seminal work of Lisa Randall and Raman Sundrum$^8$, consistency can be achieved by replacing their parameters $r_c^2$ and $\Lambda$ by $- r_c^2$ and $-\Lambda$, respectively.  Here  a new spinorial theory of physics is developed, built on Einstein's general relativity$^3$ and using the unifying triality concept of Elie Cartan$^{9, 10}$: the triality links space-time with two twistor spaces$^{[9-14]}$. Unification entails that space-time acquires two extra dimensions, each of which must be time-like. The experimental device known as the Large Hadron Collider, which is just now coming online, is expected to put the higher-dimensional theories and this prediction in particular to the test$^{15,16}$.} 
\end{quote}
\end{center}
\eject\noindent
\noindent The present theory has three key features:
\begin{itemize} \item  A powerful new spinor transform is constructed in general relativity, the $\Xi$-transform, solving a forty-year old problem posed by Roger Penrose $^{[11-14]}$: to find a non-local, essentially spinorial approach to fundamental physics.  
\item  It gives a co-ordinate free definition of chaos for space-times. Typical definitions of chaos use a preferred time co-ordinate, for example to define the Lyapunov exponents$^{17}$, violating the basic principles of general relativity.
\item It rounds out the "primordial theory" of the author and Philip Tillman$^{9, 18, 29}$, which supposes that there is a triality symmetry of the type developed by Elie Cartan$^{10}$, associated with the real Lie group $\mathbb{O}(4, 4)$.   In particular triality requires that space-time extends minimally to six-dimensions, of signature $(3, 3)$, so it predicts \emph{two extra timelike dimensions}.
\end{itemize}
The triality links  real vector spaces $\mathbb{A}$, $\mathbb{B}$ and $\mathbb{C}$, each of dimension eight and each with a dot product of signature $(4, 4)$, by a real-valued trilinear form, $(xyz)$, for $x$ in $\mathbb{A}$, $y$ in $\mathbb{B}$ and $z$ in $\mathbb{C}$$^{9, 10, 18}$. Dualizing gives maps $\mathbb{A}\times \mathbb{B} \rightarrow \mathbb{C}$,  $\mathbb{B}\times \mathbb{C}  \rightarrow \mathbb{A}$ and $\mathbb{C}\times \mathbb{A}\rightarrow \mathbb{B}$, denoted by parentheses$^{9, 18}$.  Then, for example, $((xy)x) = x.x\hspace{2pt} y$, and $(xy).z = (zx).y = (yz).x = (xyz)$, for any $x$, $y$ and $z$ in $\mathbb{A}$, $\mathbb{B}$ and $\mathbb{C}$$^{9, 18}$. \\\\Fix a null vector $y \ne 0$ in $\mathbb{B}$.  The set $\mathcal{N}_y$, of all $x$ in $\mathbb{A}$, such that $(xy) = 0$ is a totally null, self-dual, four-dimensional subspace of $\mathbb{A}$.   The restriction to $\mathcal{N}_y$ of the three-form $\omega_x = x\wedge dx \wedge dx \wedge dx$, factors: $\omega_x = \tau(y \otimes y) \alpha_x$, where $\tau$ is a canonical isomorphism of the space of trace-free symmetric elements of $\mathbb{B}\otimes \mathbb{B}$ with the space of self-dual elements of the fourth exterior product of $\mathbb{A}$ with itself (each space has dimension $35$) and $\alpha_x$ is an ordinary three-form, homogeneous of degrees minus four and minus two in the variables $x$ and $y$, respectively.  For integral $k$, denote by $\mathcal{H}(\mathbb{A}, k)$ the space of smooth functions $f(x)$, defined for non-zero null vectors $x$ in $\mathbb{A}$, that are homogeneous of degree $k$: $f(tx) = t^k f(x)$, for any non-zero real $t$.   Define $\mathcal{H}(\mathbb{B}, k)$ and  $\mathcal{H}(\mathbb{C}, k)$ analogously.  For $f(x) \in \mathcal{H}(\mathbb{A}, -4)$, the following integral is well-defined:
\[ \Xi_\mathbb{A}^\mathbb{B}(f)(y) = \int_{\mathcal{N}_y - \{0\}}  f(x) \alpha_x.\]
The integral is taken over a oriented three-sphere in the space $\mathcal{N}_y - \{0\}$, surrounding the origin.
The output function  $\Xi_\mathbb{A}^\mathbb{B}(f)$ belongs to $\mathcal{H}(\mathbb{B}, -2) $, so $\Xi_\mathbb{A}^\mathbb{B}$ gives a \emph{ natural integral transformation}: $ \displaystyle{\Xi_\mathbb{A}^\mathbb{B}:  \mathcal{H}(\mathbb{A}, -4) \rightarrow  \mathcal{H}(\mathbb{B}, -2)}$, the $\Xi$-transform.  There are \emph{six} such transforms, one for each ordered pair from the set $\{ \mathbb{A}, \mathbb{B}, \mathbb{C}\}$. 
\eject\noindent Now the work of Robin Graham, Ralph Jenne, Lionel Mason and the author$^{19}$ yields a canonical conformally invariant second order differential operator, $\square_{\mathbb{A}}$, mapping $  \mathcal{H}(\mathbb{A}, -2)$ to $ \mathcal{H}(\mathbb{A}, -4)$, with analogous operators $\square_{\mathbb{B}}$ and $\square_{\mathbb{C}}$  for the spaces $\mathbb{B}$ and $\mathbb{C}$, respectively.  The first major result is:
\[ \square_{\mathbb{B}} \circ \Xi_\mathbb{A}^\mathbb{B} = \Xi_\mathbb{A}^\mathbb{B}  \circ \square_{\mathbb{A}} = 0.\]    
The \emph{physical interpretation} is that the space of null rays in $\mathbb{A}$ is a six-dimensional generalization of Minkowski spacetime (equipped with a conformal structure of signature $(3, 3)$), into which the ordinary Minkowksi spacetime naturally embeds (see below), whereas the null rays in $\mathbb{B}$ and $\mathbb{C}$ are six-dimensional spaces of null twistors, whose real scaling is factored out$^{[11-14]}$.  Then the $\Xi$-transform gives a \emph{six-fold binding together} of these spaces, transferring information between the spaces, with \emph{losses} quantified by the wave operator.\\\\
Shockingly,  this transform generalizes to an \emph{arbitrary} four-dimensional space-time $\mathbb{M}$, with metric $g$.  Let $\epsilon$ denote the complex spinor symplectic form of $\mathbb{M}$, so that $g = \epsilon \otimes \overline{\epsilon}$ (the bar denotes complex conjugation)$^{13}$. Denote by $\mathbb{S}^*$ the co-spinor bundle of all pairs $(x, \pi)$ with $x$ in $\mathbb{M}$ and $\pi$ a non-zero co-spinor at $x$. Each such co-spinor $\pi$ naturally represents a future pointing non-zero null co-vector denoted $p_\pi$, such that $p_{t\pi} = |t|^2 p_\pi$, for any complex number $t \ne 0$$^{13}$.  The null geodesic spray $\mathcal{N}$ is the vector field on $\mathbb{S}^*$, whose integral curves represent a null geodesic with a parallely propagated spinor $\pi \ne 0$, such that $g^{-1}(p_\pi)$ is tangent to the null geodesic.   For integral $k$, denote by $\mathcal{T}(k)$ the space of all \emph{twistor} functions $f(x, \pi)$ of degree $k$:  so  $\mathcal{N}(f) = 0$ and $f(x, t\pi) = t^{k} f(x, \pi)$, for any \emph{real} number $t\ne 0$$^{13}$.  Note that the elements of $\mathcal{T}(k)$ depend on six free real variables.  \\\\
For $f(x, \pi)$ in $\mathcal{T}(-4)$, the general $\Xi$-transform is given by the integral formula:
\[ \Xi(f)(\gamma(\eta)) =  i \int_{\Gamma_{\gamma(\eta)}}   f(x, \pi) \epsilon^{-1} (\pi, d\pi)\wedge  \overline{\epsilon}^{-1} (\overline{\pi}, d\overline{\pi})\wedge \theta.p_\pi.\]
Here $\theta$ is the vector-valued canonical one-form of $\mathbb{M}$, pulled back to the co-spin-bundle and the dot represents the canonical pairing of a co-vector and a vector.  Also $d\pi$ is the tautological co-spinor-valued one-form on the co-spin-bundle representing the Levi-Civita spin connection.  Next $\gamma(\eta)$ is a future-pointing null geodesic in $\mathbb{M}$ with parallely propagated non-zero spinor $\eta$, such that  $g^{-1}(p_\eta)$ is tangent to the null geodesic.  Finally $\Gamma_{\gamma(\eta)}$ is a "fattened" null geodesic: the three-manifold consisting of all triples $(x, \pi, \eta)$ with $\pi$ and $\eta$ co-spinors at $x$, such that $x$ lies in $\gamma(\eta)$ and $\{ \pi, \eta\}$ is a \emph{normalized} spin-frame: $\pi\otimes \eta - \eta\otimes \pi = \epsilon$.
\eject\noindent 
It is easily checked that $\Xi(f) \in \mathcal{T}(-2)$, so we have a \emph{natural non-local spinor transform}, called the $\Xi$-transform from $\mathcal{T}(-4)$ to $\mathcal{T}(-2)$.    
This transform then is the long-sought answer to Penrose's challenge.  It has the following properties:
\begin{itemize} \item It is conformally invariant.
\item It uses the \emph{phase} of the spinors in an essential way:  the space of null geodesics in space-time is five-real dimensional; the sixth variable in our theory is the spinor phase.
\item \emph{Mutatis mutandis}, for the case of conformally flat space-time, it is equivalent to the $\Xi_\mathbb{B}^\mathbb{C}$ transform described above.
\item In conformally flat space-time, there is a natural conformally invariant operator, denoted $\square$, mapping $\mathcal{T}(-2)$ to $\mathcal{T}(-4)$, such that $\square \circ \Xi = \Xi\circ \square = 0$.
\item It completes the "primordial theory" providing the analytical component to accompany the previous geometric and algebraic constructions and showing how to extend that theory to curved space-time$^{18, 29, 30}$.
\end{itemize} 
We would like to understand the meaning of the $\Xi$-transform.  Of principal interest is the nature of its image.  On the basis of examination of the transform in various prototypical space-times, particularly those of Devendra Kapadia and the author$^{20}$, we are led to the following definition: 
\begin{itemize} \item  The space-time is \emph{coherent} if and only if the image of the $\Xi$-transform obeys a \emph{pseudo-differential equation}; if not the space-time is said to be \emph{chaotic}. 
\end{itemize} 
We view the coherence condition as a resonance or tuning of the space-time; in particular it is \emph{non-perturbative} in character.  We conjecture that all space-times that have previously been treated by twistor methods are coherent: for example, the stationary axi-symmetric space-times, as analyzed by Richard Ward$^{21}$.   These include the fundamental space-times of physical interest: the solutions of Karl Schwarzschild$^{22}$ and Roy Kerr$^{23}$.  Similarly we conjecture that the space-times of Vladimir Belinski, Isaak Khalatnikov and Evgeny Lifshitz$^{24}$ are chaotic in our sense.
\eject\noindent
The triality spaces have conformal symmetry group the group $\mathbb{O}(4, 4)$.   We need to reduce this to $\mathbb{O}(2, 4)$ to give the conformal symmetries of Minkowksi space-time.  Also, for  the two twistor spaces, we need to reduce to $\mathbb{U}(2, 2)$, to recover the standard successful quantum twistor description of massless particles in terms of sheaf cohomology, due to Lane Hughston, Penrose and the author$^{11,13}$. Surprisingly, we can achieve these reductions with a common mechanism:  a conformal symmetry of rotational type, whose orbits are circles about a four-dimensional invariant "axis": the conformally compactified Minkowki space-time. We may write the equation of the null cone of $\mathbb{A}$ as $q\overline{q} = r\overline{r}$, where $q$ and $r$ are non-zero quaternions, and the bar denotes the conjugation of quaternions.  Then the rotations are $(q, r) \rightarrow (q, e^{it} r e^{-it})$, for $t$ real, where $i \ne 0$ is a unit imaginary quaternion.  The axis comprises all pairs $(q, r)$ with $r = u + iv$, for $u$ and $v$ real.  Then $q\overline{q} = u^2 + v^2$, giving the correct $\mathbb{O}(2, 4)$ structure.  Simultaneously, the operator $i$ automatically gives the other triality spaces their needed complex structures, allowing the correct definitions of sheaf cohomology and massless particles.\\\\
Remarkably this same idea extends to an arbitrary space-time.   The conventional space-time structure whose information we wish to preserve is the Fefferman tensor $\mathcal{F} = i\theta^a\otimes ( \overline{\pi}_A d\pi_{A'} - \pi_{A'} d\overline{\pi}_A) $, using the abstract index formalism of Penrose$^{14}$, which is defined on the spin-bundle and which plays three roles: its skew part is the form used by Edward Witten$^{25}$ to control the space-time energy; the exterior derivative of the skew part gives rise to a form which controls the Einstein field equations$^{26}$; its symmetric part gives the \emph{central fact of twistor theory} and, in particular controls the hypersurface twistor theory$^{26, 27}$. \\\\
This structure extends naturally and beautifully to six-dimensional space-time, necessarily of signature $(3, 3)$, where the tensor is now given simply by the formula:  $\mathcal{F} = \theta^{\alpha\beta}\otimes \pi_\alpha d\pi_\beta$; here the co-spinors, $\pi_\alpha$, of real dimension four, transform according to a fundamental representation of the group $\mathbb{SL}(4, \mathbb{R})$; also the canonical one-form $\theta^{\alpha\beta}$ is skew, so has the requisite six degrees of freedom.  There are two main points. First on restriction to the original space-time submanifold, the spinors lose no information, the correspondence being $\pi_\alpha \rightarrow (\pi_{A'} , \overline{\pi}_A)$.  Second, in general, the six-dimensional spin connection $d$ would have extra terms of spin two over and above those of general relativity, on restriction to the space-time.   However these terms can be eliminated by precisely  the same mechanism that for conformally flat space allows the correct definition of massless particles: we require that there be a conformal Killing symmetry of rotational type, whose axis is the space-time.
\\\\So we predict that spacetime extends to six dimensions, of signature $(3,3 )$, with a rotational symmetry: this means that the effective structure is the anti-de-Sitter group,  allowing contact with the important work of Juan Maldacena$^{28}$; also from the viewpoint of the work of Randall and Sundrum$^{7, 8}$, space-time appears as a kind of brane or orbifold, the main difference with the philosophy of their work being the difference over signature.  Further we can now systematically go through the canon of string theory, appropriately adapting its concepts to the present situation, thereby achieving at least the outline of a synthesis for basic physics.  In particular the fundamental string amplitude, the so-called "trouser-pants" diagram$^{4}$, will become an amplitude relating three strings, one in each of three \emph{different} spaces, one being the extended space-time and the others being the two twistor spaces.     Finally the three spaces will be linked by the fundamental quantum fermionic fluid of Shou-Cheng Zhang and Jiangping Hu$^{29, 18}$, the spaces arising at the boundaries of the fluid, the excitations at the boundary giving rise to the structure of the spaces.


\newpage\begin{thebibliography}{30}
\bibitem{euc1} Euclid of Alexandria,\hspace{2pt}  \emph{Euclidus Opera Ominia}, \hspace{2pt}I.L. Heiberg \& H. Menge (editors), Teubner (1883-1916), circa 300 BC.
\bibitem{mink1} Hermann Minkowksi,\hspace{2pt}  \emph{Raum und Zeit}\hspace{2pt}  in\hspace{2pt}  \emph{  Vortr\"age von der 80. Naturforcherversammlung zu K\"oln}, \hspace{2pt} Physikalische Zeitschrift, \textbf{10}, 104-111, 1908.
\bibitem{ein1} Albert Einstein,\hspace{2pt}  \emph{Die Grundlage der allgemeinen Relativitaetstheorie, Annalen der Physik},\hspace{2pt}   \textbf{49}, 769-822, 1916.
\bibitem{wit2}Michael Green, John Schwartz and Edward Witten,\hspace{2pt} 
\emph{Superstring theory, Volume 1, Introduction}, \hspace{2pt}  Cambridge
Monographs on Mathematical Physics, Cambridge University Press,
Cambridge, 1987.
\bibitem{hamed1} Nima Arkani-Hamed, Savas Dimopoulos and Georgi Dvali, \hspace{2pt} \emph{The hierarchy problem and new dimensions at a millimeter}, \hspace{2pt} Physics Letters, \textbf{B 429}, 263-272, 1998.
\bibitem{dien1}Keith Dienes, Emilian Dudas and Tony Gherghetta, \hspace{2pt} \emph{Extra spacetime dimensions and unification}, \hspace{2pt} Physics Letters, \textbf{B 436}, 55-65, 1998.
\bibitem{rs1} Lisa Randall and Raman Sundrum, \hspace{2pt} \emph{An Alternative to Compactification}, \hspace{2pt}Physical Review Letters, \textbf{83}, 4690-4693, 1999.
\bibitem{rs2} Lisa Randall and Raman Sundrum, \hspace{2pt} \emph{A Large Mass Hierarchy from a Small Extra Dimension}, \hspace{2pt}Physical Review Letters, \textbf{83}, 3370-3373, 1999.
\bibitem{ba1}John Baez,\hspace{2pt} \emph{The Octonions},\hspace{2pt}Bulletin of the American Mathematical Society, \textbf{39}, 145-205, 2002.
\bibitem{ca1}Elie Cartan, \hspace{2pt} \emph{Le principe de dualit\'e et la th\'eorie des groupes simple et semi-simples},\hspace{2pt} Bulletin de  Science Math\'ematique,
\textbf{49}, 361-374, 1925.
\bibitem{pen0} Roger Penrose,\hspace{2pt} \emph{The road to reality: a complete guide to the laws of the universe},\hspace{2pt} Alfred A. Knopf, New York, 2005.
\bibitem{pen3} Roger Penrose, \hspace{2pt}\emph{Twistor theory: its aims and
achievements} in \hspace{2pt} \emph{Quantum gravity: An Oxford Symposium}, \hspace{2pt}editors Christopher Isham, Roger Penrose and Dennis Sciama, Oxford: Clarendon Press, 1975.
\bibitem{pen5}Roger Penrose and Wolfgang Rindler, \hspace{2pt}\emph{Spinors and
Space-Time.  Volume 2: Spinor and Twistor Methods in Space-Time
Geometry},\hspace{2pt} Cambridge: Cambridge University Press, 1986.
\bibitem{pen4} Roger Penrose and Wolfgang Rindler, \hspace{2pt}\emph{Spinors and
Space-Time.  Volume 1: Two-spinor Calculus and Relativistic Fields},\hspace{2pt}
Cambridge: Cambridge University Press, 1984.
\bibitem{all1} Benjamin Allanach, Kosuke Odagiri, Matt Palmer, M. Andy Parker, Ali Sabetfakhri, BryanWebber, \hspace{2pt}\emph{Exploring Small Extra Dimensions at the Large Hadron Collider}, \hspace{2pt}Journal of High Energy Physics, \textbf{0212}, 039-060, 2002.
\bibitem{bens1} Kamal Benslama, \hspace{2pt} \emph{Search for extra dimensions with ATLAS at LHC}, \hspace{2pt} in \emph{
Fundamental Interactions: Proceedings of the Nineteenth Lake Louise Winter Institute - Lake Louise, Alberta, Canada, 15 - 21 February 2004},\hspace{2pt}  editors Alan Astbury, Bruce Campbell, Faqir Khanna, Manuella Vincter, World Scientific Publishing Company, 2005.
\bibitem{cvi1} Predrag Cvitanovi\'c, Roberto Artuso, Per Dahlqvist, Ronnie Mainieri, G\'abor Vattay, Niall Whelan and Andreas Wirzba, \hspace{2pt}\emph{Chaos: Classical and Quantum},\hspace{2pt} Niels Bohr Institute, Copenhagen, 2004.
\bibitem{moab11} George Sparling and Philip Tillman, \hspace{2pt} \emph{A primordial theory}, \hspace{2pt} cond-mat/0401015, 55 pages, submitted to the Journal of Mathematical Physics, October 2006.
\bibitem{graham1} C. Robin Graham, Ralph Jenne, Lionel Mason and George Sparling,  \hspace{2pt} \emph{Conformally invariant powers of the Laplacian}, \hspace{2pt}Journal of the London Mathematical Society, \textbf{46(3)}, 557-565, 1992. 
\bibitem{moab3}Devendra Kapadia  and George Sparling, \hspace{2pt}\emph{A class of conformally Einstein metrics}, \hspace{2pt}Classical and Quantum
Gravity, \textbf{24}, 4765-4776, 2000.
\bibitem{ward1} Richard Ward,\hspace{2pt} \emph{Stationary axisymmetric space-times: a new approach}, \hspace{2pt}General Relativity and Gravitation, \textbf{15}, 105 (1983).
\bibitem{schw1} Karl Schwarzschild, \emph{\"Uber das Gravitationsfeld eines Massenpunktes nach der Einsteinschen Theorie}, \hspace{2pt}Sitzungsberichte der Kšniglich Preussischen Akademie der Wissenschaften \textbf{1}, 189-196, 1916.
\bibitem{kerr1} Roy Kerr, \emph{Gravitational Field of a Spinning Mass as an Example of Algebraically Special Metrics}, Physical  Review  Letters \textbf{11}, 237-238, 1963.
\bibitem{bkl1} Vladimir Belinski, Isaak Khalatnikov and Evgeny Lifshitz,\hspace{2pt}  \emph{Oscillatory approach to a singular point in the relativistic cosmology}, \hspace{2pt}Advances in Physics, \textbf{19}, 525-573, 1970.
\bibitem{witten1} Edward Witten,\hspace{2pt} \emph{A new proof of the positive energy theorem},\hspace{2pt} Communications in Mathematical Physics 80, 381-402, 1981.
\bibitem{moab2} George Sparling, \hspace{2pt} \emph{The twistor theory of hypersurfaces in
space-time}, \hspace{2pt}  in\hspace{2pt}  \emph{Further Advances in Twistor Theory,
Volume III},\hspace{2pt}  editors Lionel Mason, Lane Hughston,  Piotr Kobak and Klaus Pulverer, London: Pitman
Press, 2001.
\bibitem{feff1} Charles Fefferman, \hspace{2pt}\emph{The Bergman kernel and biholomorphic mappings of pseudoconvex domains}, \hspace{2pt}Inventiones Mathematicae,  \textbf{26}, 1Ð65, 1974. 
\bibitem{mal1} Juan Maldacena, \hspace{2pt}\emph{The Large N Limit of Superconformal Field Theories and Supergravity}, \hspace{2pt} Advances in Theoretical Mathematical Physics \textbf{2}, 231-252, 1997.
\bibitem{zh1}Shou-Cheng Zhang and Jiangping Hu, \hspace{2pt}\emph{A Four-Dimensional Generalization of the Quantum Hall Effect},\hspace{2pt} Science,
\textbf{294}(5543), 823-828, 2001. 
\end{thebibliography}
\end{document}